\documentclass[twocolumn,10pt,aps,nofootinbib]{revtex4-1}

\def\mysections#1{{\bf #1.} }

\usepackage[dvips]{graphicx}
\usepackage{epsfig,amsmath,amssymb,verbatim,mathrsfs,array,layout,textcomp,latexsym, slashed,graphicx,bbm}

\usepackage[normalem]{ulem}
\usepackage{appendix}

\usepackage{relsize}

\usepackage{rotating}
\usepackage[hidelinks]{hyperref}
\usepackage[usenames,dvipsnames]{color}

\usepackage{soul}

\newcommand{\be}{\begin{equation}}
\newcommand{\ee}{\end{equation}}
\newcommand{\bea}{\begin{eqnarray}}
\newcommand{\eea}{\end{eqnarray}}

\definecolor{gbcolor}{rgb}{.1,.2,.8}
\definecolor{gbcolor2}{rgb}{.8,.1,.7}

\def\beq{\begin{equation}}
\def\eeq{\end{equation}}

\newcommand{\eq}[1]{(\ref{#1})}

\begin{document}

\begin{flushleft}                          
\footnotesize {IFT-UAM/CSIC-19-87}
\end{flushleft}

\title{X-ray and gamma-ray limits on the \\primordial black hole abundance from Hawking radiation}

\author{Guillermo Ballesteros$^{1,2}$}\email{guillermo.ballesteros@uam.es}
\author{Javier Coronado-Bl\'azquez$^{1,2}$}\email{javier.coronado@uam.es}
\author{Daniele Gaggero$^{1}$}\email{daniele.gaggero@uam.es}
\affiliation{$^1$Instituto de F\'isica Te\'orica UAM/CSIC,
Calle Nicol\'as Cabrera 13-15, Cantoblanco E-28049 Madrid, Spain}
\affiliation{$^2$Departamento de F\'isica Te\'orica, Universidad Aut\'onoma de Madrid (UAM)  Campus de Cantoblanco, 28049 Madrid, Spain}

\begin{abstract}

The non-observation of extragalactic Hawking radiation from primordial black holes of $10^{16}$g sets a conservative strong bound on their cosmological abundance.  We revisit this bound and show how it can be improved (both in mass reach and strength) by an adequate modeling of the combined AGN and blazar emission in the MeV range. We also estimate the sensitivity to the primordial black hole abundance of a future X-ray experiment capable of identifying a significantly larger number of astrophysical sources contributing to the diffuse background in this energy range.

\end{abstract}

\maketitle

\section{Introduction}

The isotropic background radiation that fills the Universe and extends over more than 16 orders of magnitude in frequency --from radio waves all the way up to high-energy gamma rays \cite{Hill:2018trh}--  carries information on the emission mechanisms of different astrophysical and cosmological sources over the history of the Universe, and can possibly shed further light on the nature of the elusive dark matter (DM) that constitutes the largest fraction of its mass. 

We consider the hypothesis that the bulk of the DM is made up of {(non-rotating)} black holes of primordial origin (PBHs), formed from the collapse of overdense Hubble patches prior to the big-bang nucleosynthesis epoch \cite{ZelNov,Hawking:1971ei}. We focus on the current PBH mass window for DM ranging from approximately  $10^{17}$~g to $10^{19}$~g, revisiting the Hawking radiation constraints on their abundance from extragalactic gamma-ray data \cite{Carr:2009jm} and showing how future gamma- and X-ray observations\footnote{The frontier between gamma- and X-rays is not sharply defined. A reasonable distinction can be made setting it at $\sim 100\, \rm{keV}$.} with an increased sensitivity have the potential for discovering a population of PBHs comprising the totality of the DM.  

This mass range is indeed particulary relevant for DM, given that it has been found that previously claimed femtolensing bounds \cite{Barnacka:2012bm} were marred by an inadequate treatment of the involved optics, leaving much of that window open  \cite{Katz:2018zrn}. In addition,  the $\sim 10\%$ limit on the abundance of PBHs at $10^{19}$~g -- $10^{20}$~g  from the observed distribution of white dwarfs \cite{Graham:2015apa}, as well as from the disruption of neutron stars in the PBH {mass} range from $10^{19}$~g to $10^{23}$~g \cite{Capela:2013yf}  has been challenged \cite{Montero-Camacho:2019jte}, opening the possibility that PBHs of mass below
$\sim 5\times 10^{22}$ g --with higher masses being constrained by microlensing \cite{Niikura:2017zjd}-- could explain all the DM. 
 The lower end of the current PBH mass window for DM, at $\sim 10^{17}$ g, comes instead from Hawking evaporation limits. As we already mentioned, in this work we focus specifically on extragalactic gamma-ray bounds from evaporation \cite{Carr:2009jm}. However, there are other phenomena, related to Hawking radiation, which have been used to constrain this low mass region: the Voyager measurements of $e^\pm$ \cite{Boudaud:2018hqb}, the 511 keV positron-electron  annihiliation line from INTEGRAL \cite{DeRocco:2019fjq,Laha:2019ssq,Dasgupta:2019cae}, the non-detection of a neutrino flux from PBH emission at Superkamiokande \cite{Dasgupta:2019cae}, distortions on the CMB anisotropies  \cite{Poulin:2016anj,Stocker:2018avm}, and the Galactic emission of gamma/X-rays \cite{Carr:2016hva} (see also \cite{Laha:2020ivk}). The advantage of the bounds coming from the possible PBH extragalatic  emission is that they are free from Galactic propagation uncertainties. As we will see,  making use of their full potential requires an adequate characterization of the emission from other astrophysical sources, mostly active galactic nuclei (AGNs) and  blazars.

Hawking radiation \cite{Hawking:1974sw,Hawking:1974rv,Page:1976wx, MacGibbon:1990zk, MacGibbon:1991vc} is an approximately thermal particle emission expected to be emitted by black holes, with temperature $T=(8\pi k_B)^{-1} c^2\, m_P^2/M\simeq6\times10^{-8} M_\odot/M$ K, being $M$ the mass of the black hole, $M_\odot=3\times 10^{33}$ g the mass of the Sun and $m_P=\sqrt{{\hbar\,c}/{G}}$ the Planck Mass. In Ref.\ \cite{Carr:2009jm}, a conservative (but nonetheless stringent) upper bound on the cosmological PBH abundance was set for $M \lesssim 10^{17}$ g by comparing the predicted Hawking gamma-ray emission with the isotropic gamma-ray background in the approximate energy range 0.1 MeV --- 10 GeV that was measured by EGRET \cite{Strong:2004ry}, Fermi-LAT \cite{2010PhRvL.104j1101A} and COMPTEL \cite{Weidenspointner:2000aq}. This bound has recently been updated in \cite{Arbey:2019vqx} (with the data from the same experiments), finding a good agreement with \cite{Carr:2009jm}. 

The main focus of our analysis is on the isotropic gamma- and X-ray background in the 10 keV -- MeV domain. Using a power-law modeling of such background --motivated by the assumption that a population of unresolved extra-Galactic sources (mainly AGNs and blazars) represent the main contribution to it-- we place an upper limit on the abundance of PBHs (as a function of their mass) by considering an array of datasets acquired by a variety of missions has been proposed over the latest decades. We also estimate the expected improvement in the bounds from a putative future experiment. We do so by assuming  that such a (more sensitive) experiment will resolve a significantly larger number of individual AGNs and blazars and will therefore provide a lower isotropic unresolved background for energies above $\sim 200$ keV. We show that, under this assumption, a significantly better upper limit on the PBH abundance than the current one may be placed in the future. This result motivates the investment in future gamma- and X-ray experiments in this range, as well as in further theoretical studies geared towards a more precise modeling of astrophysical sources.

\section{Hawking radiation from PBHs}
\label{sec:Hawking}

The photon emission from a population of PBHs of mass $M$ accounting for a fraction $f=\Omega_{PBH}/\Omega_{DM}$ of the total DM density in the Universe is
\begin{align}
\Phi_M=\frac{{\rm d} N}{{\rm d} E\,{\rm d} t} =  f\frac{c\,\rho}{4\pi\,M} \int { {\rm d}z \frac{e^{-\tau(z)}  }{ H(z)} \, \Psi_M[(1+z)E]} \,,
\label{eq:x-ray}
\end{align}
where $\rho=2.17 \times 10^{-30} {\rm g}/{\rm cm}^3$ is the current DM density of the Universe \cite{Aghanim:2018eyx}, and $H(z)$ is the Hubble rate of expansion as a function of redshift. The function $\Psi_M[E]$ denotes the differential flux emitted by a single PBH, as a function of the energy $E$, per unit of energy and time. For PBHs of masses above $10^{16}$ g  is well approximated by the primary\footnote{The spectrum of BHs may feature a secondary emission component, depending on their mass, which is due to the interactions among the primary emitted particles, see e.g.\ \cite{Carr:2009jm}.} Hawking emission:
\begin{align}
\Psi_M[E]=(2\pi\hbar)^{-1}\Gamma_s/(\exp(E/k_B T)-1)\,,
\end{align}
where the so-called grey factor $\Gamma_s$ is a function of $M$ and $E$. In the high-energy limit  $E\gg k_B T$, the grey factor approximately satisfies $\Gamma_s 
\propto (M/m_P)^2 (E/ m_P\, c^2)^2$; whereas for $E\ll k_B T$, $\Gamma_s 
\propto (M/m_P)^4 (E/ m_P\, c^2)^4$ \cite{MacGibbon:1990zk}. These expressions are insufficient to render adequately the peak height and position of $\Psi_M[E]$, which is best computed numerically. To do so we use the public code BlackHawk \cite{Arbey:2019mbc}, which also allows to include the (subdominant) secondary emission. We find that the differential flux for BHs of mass between $10^{16}$g and $10^{20}$g can be approximated by
\begin{align}
\Psi_M[E]\simeq \frac{2.5\times 10^{21}\,{\rm GeV}^{-1}{\rm s}^{-1}}{(M_{18}\,E/E_0)^{-2.7}+(M_{18}\,E/E_0)^{6.7}}\,,
\end{align}
where $E_0=6.54\times10^{-5}$ GeV and $M_{18}\equiv M/10^{18}\rm g$. This approximation is accurate to better than $\sim 1\%$ around the emission's peak (until $\Psi_M[E]$ decreases an order of magnitude), which is enough for our purposes. Nevertheless, we obtain the bounds on the PBH abundance from the instantaneous spectra given by BlackHawk. 

The factor $(1+z)$ inside $\Psi_M[(1+z)E]$ accounts for the Doppler shift from the time of emission to the time of arrival to the detector. The optical depth $\tau(z)$ describes the attenuation due to the propagation of the signal over the relevant cosmological redshifts. Unlike for hard gamma rays, this is negligible for soft gamma-rays and $X$-rays. The integrand in \eq{eq:x-ray} decreases very rapidly with $z$ and accurate results are obtained integrating up to $z\sim \mathcal{O}(100)$.

\begin{figure}[t]
\includegraphics[width=0.5\textwidth]{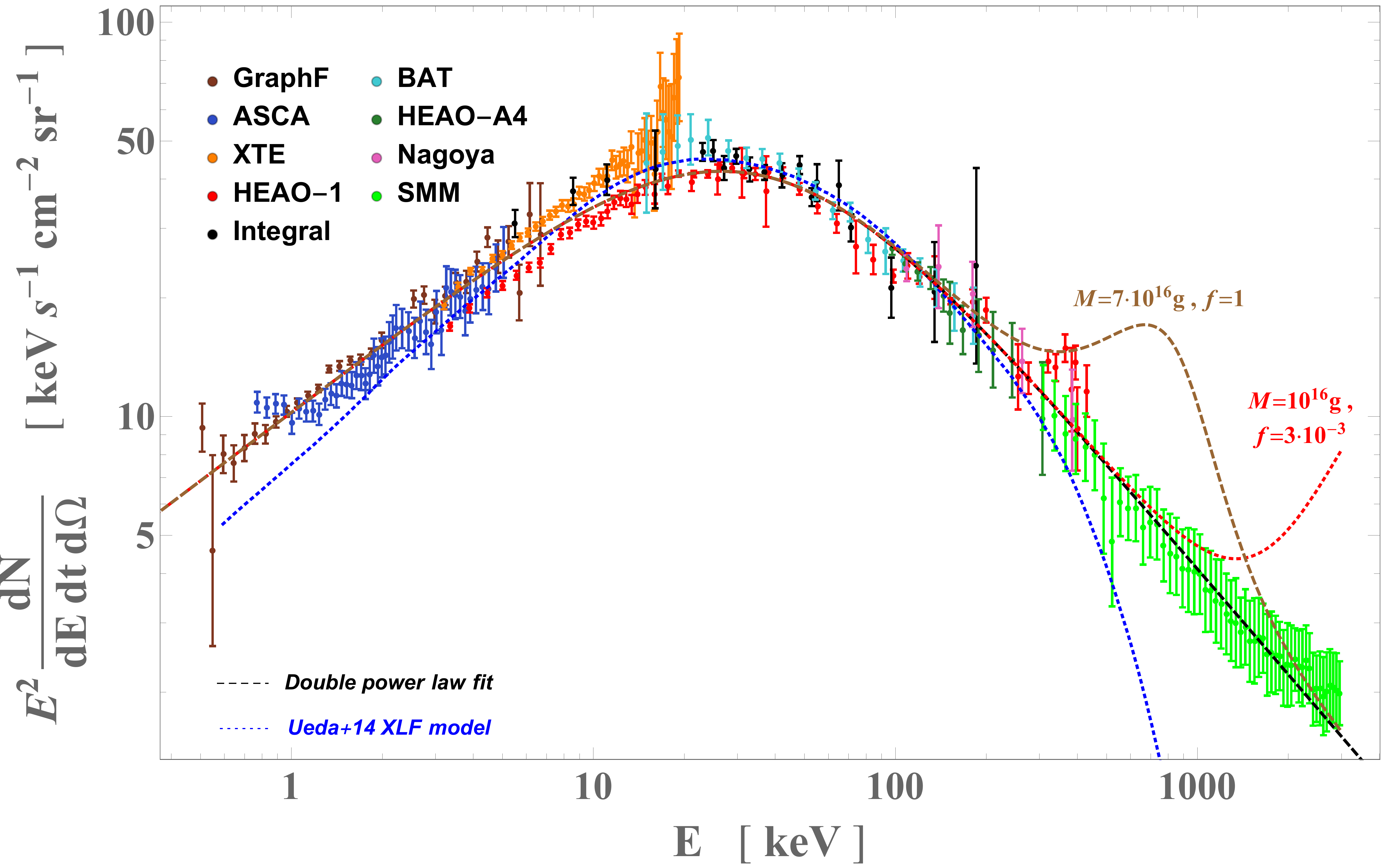}
\caption{Cosmic X-ray background spectrum, as measured by various experiments. Overimposed are the Ueda+14 model (blue dashed line), the fit to a double power-law (black dashed line) of Eq.\ \ref{eq:x-raym}, and the corrections to the latter due to two hypothetical monochromatic PBH distributions with different masses $M$ and cosmological abundances $f=\Omega_{PBH}/\Omega_{DM}$.}
\label{fig:xray_plot}
\end{figure}

\section{The X-ray and gamma-ray AGN background}
\label{sec:XrayBackground}

There has been a considerable effort dedicated to interpreting the measurements of the X-ray and gamma-ray background from keV energies all the way up to $\sim 100$~GeV in terms of a superposition of a large number of unresolved extra-Galactic sources. In particular, the data in the range $\sim 5$--200 keV  observed by  \textit{Swift}/BAT \cite{Swift_paper}, MAXI \cite{MAXI_paper}, ASCA \cite{ASCA_paper}, XMM-Newton \cite{XMM_paper}, Chandra \cite{Chandra_paper} and ROSAT \cite{ROSAT_paper} are well reproduced by a population synthesis model of active galactic nuclei (AGNs) developed by Ueda et al.\ in \cite{Ueda:2014tma} (see the blue dotted line in Figure \ref{fig:xray_plot}).  
AGNs are powered by gas accretion onto a supermassive black hole and are very efficient X-ray emitters. The model of \cite{Ueda:2014tma} is based on the extrapolation of the luminosity functions of AGNs in different redshift ranges inferred by a sample of 4039 AGNs in soft (up to 2 keV) and/or hard X-ray bands ($>$2 keV). The objects in the sample include both Compton-thin and Compton-thick AGNs (with the latter being heavily obscured by dust). As can be seen if Figure \ref{fig:xray_plot}, this AGN modeling fails to describe adequately the SMM data. 

Indeed, for energies above $\sim 50-100$ keV the contribution from blazars is expected to become progressively more important. These objects correspond to the
AGNs that are detected at a small angle between the accretion disk axis and the observer line of sight \cite{Urry:1995mg}; together with star-forming galaxies and radio galaxies, they are thought to dominate the GeV-TeV gamma-ray isotropic background measured by Fermi-LAT {(not shown in Figure \ref{fig:xray_plot}).}

Although the details of the intermediate MeV -- GeV  domain are still not clearly understood (see in particular the discussion in \cite{DeAngelis:2017gra}), the previous considerations lead us to employ the working assumption that a combination of different classes of extra-Galactic emitters explain the X-ray and gamma-ray unresolved diffuse background in a wide energy range  and that the emission from these populations of sources can be modeled as a superposition of featureless power-laws. 

Therefore, in the approximate energy range going from $20$~keV to $3000$~keV, which corresponds to the region where the Hawking emission from BHs in the mass range $10^{16}$~g -- $10^{19}$~g can contribute importantly to the Universe's diffuse spectrum,  we model the astrophysical background as a double power-law fit to the data from the SMM \cite{Watanabe1997}, Nagoya balloon \cite{Fukada1975}, HEAO--1 and HEAO-A4 \cite{Gruber1999,Kinzer1997} experiments. Concretely, we use the following proxy for the combined AGN and blazar emission: 
\begin{align}
\Phi_{\rm AGN}=\frac{{\rm d} N}{{\rm d} E\,{\rm d} t} \,= \frac{A}{\left(E/E_b\right)^{n_1}+\left(E/E_b\right)^{n_2}}\,.
\label{eq:x-raym}
\end{align}
For instance, assuming zero contribution to the data from PBH evaporation, the best fit is: $E_b = 35.6966$~keV, $A=0.0642\, {\rm keV^{-1}s^{-1}cm^{-2}sr^{-1} }$, $n_1=1.4199$ and $n_2=2.8956$;
see the black dashed line in Figure \ref{fig:xray_plot}.

\section{Constraints and prospects on $f$}
\label{sec:constraints}

We present now our results on the current upper limits on the  PBH abundance $f=	 \Omega_{\rm PBH}/\Omega_{\rm DM}$, in the mass window of interest, and estimate the prospects for a future MeV mission. {We recall that we assume a population of non-rotating PBHs. Angular momentum makes black holes evaporate faster, making the bounds on their abundance stronger \cite{Arbey:2019vqx}. For PBHs formed during radiation domination --which is the most common scenario-- the assumption of negligible angular momentum is the most reasonable one. Some amount of angular momentum may be expected if the PBHs form during a phase of early matter domination \cite{Harada:2017fjm}. A moderate level of rotation would not change our results significantly. Only nearly extremal black holes (which is unlikely in common PBH formation models) would lead to a significant difference. In this sense, our bounds and forecast should be regarded as conservative.}\newline

\begin{figure}[t]
\includegraphics[width=0.5\textwidth]{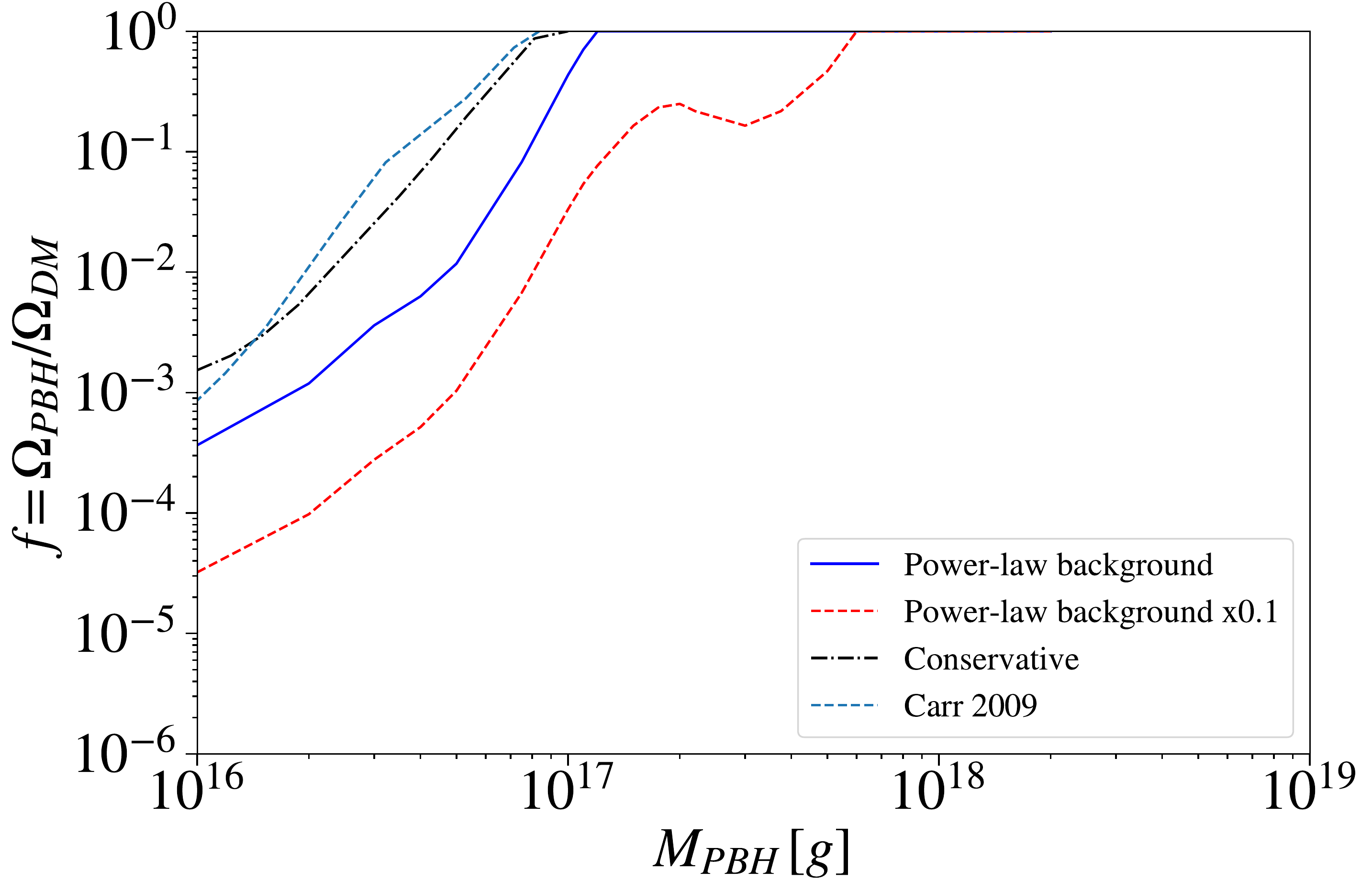}
\caption{95\% c.l.\ upper limits on the PBH abundance versus the PBH mass assuming a monochromatic mass population. The black dot-dashed line is the result of the present work neglecting the background contribution from AGNs. The light blue-dotted line is the analogous result of \cite{Carr:2009jm}, see also \cite{Arbey:2019mbc}. The blue continuous line is the bound obtained assuming a power-law modeling of the AGN and blazar contribution to the observed spectrum. The red dashed curve indicates the sensitivity achievable with an experiment capable of reducing the astrophysical background by a factor of 10}
\label{fig:mainResult}
\end{figure}

\noindent $\star$  {\bf Current bound: Conservative approach.} 

We start by deriving the present bound on $f$ under the most conservative approach, i.e.\ without assuming any astrophysical modeling of the data, as done in \cite{Carr:2009jm} (see also \cite{Arbey:2019vqx}). Under the assumption that the data are Gaussian distributed, we compute the estimator
\begin{align} \label{est}
\hat\chi^2 =\sum(D - E^2\Phi_M)^2/{\Delta^2}
\end{align}
over the energy bins for which the PBH emission $\Phi_M$ is larger than the X-ray data $D$ shown in Figure \ref{fig:xray_plot} and whose errors are denoted $\Delta$ in \eq{est}. The allowed PBH fraction $f$ for a given PBH mass $M$ at $95\%$ c.l.\ corresponds to values of $\hat\chi^2$ smaller than the threshold
\begin{align}
\hat\chi^2 \leq \chi^2_{0.05} (N-1)\,,
\end{align}
where $N$ --which depends on $f$-- is the number of bins in which the model overshoots the data. The threshold and the estimator are computed as functions of $f$ by adopting $N-1$ as the number of degrees of freedom for the $\chi^2$ distribution. The bound is shown in Figure \ref{fig:mainResult} as a black dot-dashed line. Our result is in good agreement with those of \cite{Carr:2009jm} (see blue dotted line)	 and \cite{Arbey:2019vqx} in the PBH mass region ($M \geq 10^{16}$g) that we consider, even though in the (relevant) region between 1 MeV and 20 MeV those works used COMPTEL data whereas we use SMM, which has a much smaller error and a broader range. \newline

\noindent $\star$  {\bf Sensitivity reach with the current data.} 

We assume now a double power-law fit of the AGN background akin to that of Eq.\ \eq{eq:x-raym}
and consider the following estimator, evaluated {\it over all energy bins} (of the data we have used to obtain the fit):
\begin{align}
\hat\chi^2={\sum}{\left( D - E^2 \Phi_M- E^2 \Phi_{\rm AGN} \right)^2}/{\Delta^2}\,.
\end{align}
The upper limit on the PBH abundance is set at the value of $f$ above which $\hat\chi^2$ worsens beyond the threshold for the 95$\%$ c.l.\ with respect to the minimum: i.e. the threshold is set as 
\begin{align}
\hat\chi^2 - \hat\chi^2_{min}  \leq \chi^2_{0.05} (1) \simeq 3.84\,.
\end{align}
Importantly, for each mass bin we obtain the best-fitting combination of AGN and blazar plus PBH emission by letting free the PBH fraction $f$ as well as the four parameters ($n_1$, $n_2$, $E_b$ and $A$) of Eq.\ \eq{eq:x-raym}, and then vary the parameter $f$ (recomputing the AGN and blazar contribution at each step) until the threshold above is reached.\newline

In Figure \ref{fig:mainResult} we show the results of both approaches ({\it conservative} and  {\it sensitivity reach}) for 95\% c.l.\ upper limits on $f$. Regarding the conservative approach (black solid line), our constraints are in good agreement with those from \cite{Carr:2009jm} (blue dotted line), while the bound assuming the power-law AGN model (blue solid line) is a factor $\sim10$ stronger for the same $M$ and reaches larger PBH masses (up to about $7\times10^{17}$g). \newline

\noindent $\star$ {\bf Future prospects.}

In the forthcoming {years,} both the hard X-ray and the MeV gamma-ray sky will be probed with increasing accuracy by several planned space observatories. In the MeV range, a variety of space missions have been proposed over the last years, for instance: e-ASTROGAM \cite{DeAngelis:2017gra}, AMEGO \cite{McEnery:2019tcm}, AdEPT \cite{Hunter:2013wla}, COSI \cite{Tomsick:2019wvo}, SMILE \cite{Tanimori:2015wma}.  
In the hard X-ray domain, the instrument onboard the ASTRO-H mission \cite{Matsumoto:2018mba} will measure X-rays up to $80$ keV with unprecedented accuracy.

All these instruments will detect a larger number of point sources, both AGNs and blazars. They are thus expected to characterize a lower isotropic extragalactic background due to unresolved point sources. As a consequence, a hypothetical PBH signal will be easier to detect. It is therefore important to provide a quantitative estimate of the potential of possible future experiments in either setting a stronger upper limit on the PBH abundance in the mass range under investigation, or identifying a PBH signal with sufficient significance. 

A careful assessment on how the expected increase in point-source sensitivity, especially in the MeV domain, results in a lower isotropic background due to sub-threshold sources is beyond the scope of the present work\footnote{Notice that such an analysis can be done only on a case-by-case basis, from the detailed specifications of the experiments (which in several cases are still being decided)}. Instead, we present a simple but useful calculation based on straightforward extrapolations of observed luminosity functions at low redshift. If we assume that most of the isotropic background in the MeV domain is due to blazars, and we assume a single-power law luminosity function of the form \cite{2009ApJ...699..603A}
\begin{equation} \label{spl}
N(S) =  N_0 (S/S_0)^{-\alpha}\,,
\end{equation}
where $S$ is the source flux and $\alpha$ is a positive number, the isotropic background at redshift zero due to unresolved sources is simply given by the integral:
\begin{equation}
\mathcal{I}(S_a,S_b)=\int_{S_0}^{S_{th}}  S\,\frac{dN}{dS}\,dS\,,
\end{equation}
where $S_0$ is the minimum luminosity and $S_{th}$ is the detection threshold of the experiment. If we now consider two experiments with different thresholds $S_1$ and $S_2$, the ratio between the unresolved flux in the two cases is 
\begin{equation}
r = \frac{\mathcal{I}(S_2,S_0)}{\mathcal{I}(S_1,S_0)}\,.
\end{equation}
%
In the limit of $S_0 \ll S_1,S_2$ and for a exponent $0 < \alpha < 1$ (like the ones found in \cite{2009ApJ...699..603A}),  we get: 
\begin{equation}
r \simeq  \left(\frac{S_2}{S_1}\right)^{1-\alpha}\,.
\end{equation}

Let us now consider a future experiment such as AMEGO, which is able to provide a factor of $\sim 10$ increase in point-source sensitivity with respect to COMPTEL at 1 MeV \cite{McEnery:2019tcm}. Let us assume a single power-law luminosity function for a population of blazars, as in \eq{spl}, following \cite{2009ApJ...699..603A}. For obtaining an order-of-magnitude estimate, we can ignore a possible (but mild) redshift dependence of the exponent $\alpha$. The above formalism implies that a background reduction by a factor of $10$ is possible with a luminosity function exponent $\alpha \simeq 0.5$, which is commensurate with the values quoted in \cite{2009ApJ...699..603A}.

Given these considerations, we consider here two different scenarios, characterized by {factors} of $10$ and $100$ reduction in the diffuse, unresolved background, accompanied by a reduction by the same factor of its uncertainty. We generate mock data according to this prescription, and adopt the same procedure we applied to the current data in the {\it sensitivity reach} approach described previously. The results are shown in Figure \ref{fig:mainResult}. The red dashed line indicates a potentially significant improvement of the current upper limits for a factor 10 reduction, together with a notable extension towards larger masses.

\section{Discussion and Conclusions}

In this paper we have studied the upper limits on the abundance of primordial black holes  from X-ray and gamma-ray Hawking radiation, as a function of mass and assuming a monochromatic distribution in the $10^{16}$ g -- $10^{19}$~g mass range. This approximate mass window  is particularly interesting {not only for the DM problem but also} from the model-building point of view. The known examples of the conceptually simplest mechanism capable of producing PBHs from single-field inflation --based on an approximate inflection point in the potential \cite{Ivanov:1994pa}, see e.g.\ \cite{Garcia-Bellido:2017mdw,Kannike:2017bxn,Ballesteros:2017fsr,Cicoli:2018asa,Dalianis:2018frf,Ballesteros:2019hus,Ballesteros:2020qam}-- tend to do so in this range, once a reasonable fit to the CMB data at cosmological scales is imposed, see e.g.\ the related discussion in \cite{Ballesteros:2017fsr}. As discussed in the introduction, this mass range is within the currently most promising region for the existence of a significant contribution of PBHs to the DM. 

We have first computed the upper limit with a conservative approach, by requiring that the expected signal from PBHs does not overshoot the diffuse {(extragalactic)} background. The bound we have obtained is	 mostly driven by SMM data and agrees {well} with the results presented in \cite{Carr:2009jm,Arbey:2019vqx}, which instead used COMPTEL data in the relevant energy range. This bound is competitive with the recently derived upper limits based on the Voyager $e^\pm$ data \cite{Boudaud:2018hqb} and extends to larger masses.

Then, we have considered a less conservative approach, which {takes} into account in a simple way the expected effect of known classes of unresolved astrophysical sources --AGNs and blazars in particular-- that are thought to provide the dominant contribution to the X-ray and gamma-ray diffuse isotropic background. Under the well-motivated (and common in the literature, see \cite{Ajello:2008xb}, \cite{Ueda:2014tma})  assumption  that the current data are mainly reproduced by such sources and that they can be characterized by a double power-law,  we derive a stronger upper limit on the PBH abundance based on a simultaneous fit of current data to the emission of these sources and PBHs  in the MeV domain. This (sensitivity reach) bound is about an order of magnitude stronger than the conservative upper limit described above. 

We have also considered the potential of a future, more sensitive, experiment in the MeV domain that could resolve a larger number of individual sources, therefore providing a lower and more accurate estimate of the diffuse unresolved background. The prospects, supported by the extrapolation of currently measured AGN and blazar luminosity functions, are very promising: We remark in particular that PBH masses as large as $10^{18}$ g are within reach under the assumption of a background reduction by a factor of $10$. 

The exploration of the {diffuse photon emission in the low-energy gamma-ray (or upper X-ray range)} is therefore a promising avenue towards a possible future detection of a signal associated to a population of PBHs that may constitute a significant part --perhaps even all-- {of} the DM in the Universe. 

In order to obtain further progress in this {promising PBH mass region for DM ($\gtrsim 10^{17}\rm{g}$),} a more sensitive experiment is needed {both in the MeV domain and (particularly) in the keV one,} and a more detailed understanding of the population of astrophysical sources that contribute to the bulk of the diffuse background is essential. Above 1 MeV, such modeling should potentially include not only the contributions from sources (AGNs and blazars) that are known to be dominant in the sub- and sup-Mev bands, but also possible currently subdominant sources. 

{For a fixed value of the PBH abundance $f$, the emission $\Phi_M$ grows if the mass $M$ of the PBHs is {\it decreased} --we recall that we assumed a monochromatic distribution-- and, simultaneously, the location of the peak of the PBH moves towards higher energies. This can be understood qualitatively (disregarding the secondary emission) by applying the scaling discussed in Section \ref{sec:Hawking}. As the measured diffuse background has  a decreasing overall tendency from $\sim 30\, \rm{keV}$ to $\sim 10^3\, \rm{MeV}$ (see \cite{Hill:2018trh}), the bounds on the PBH abundance become stronger than those shown in figure \ref{fig:mainResult} for smaller masses (below $10 ^{16}\,\rm{g}$). This low mass region is thus of no relevance for the DM problem and we have not considered it in this work. It is interesting though that the emission of a bright but sub-dominant population of PBHs of mass $\mathcal{O}$($10^{15}$ g - $10^{16}$ g) would peak in the 1-10 MeV region.  In this energy range, the Compton and Fermi X-ray data present a noticeable change in slope with respect to that of SMM (see e.g.\ figure 7 of ref. \cite{Hill:2018trh}). This energy region of the spectrum of the Universe is still not fully characterized from a theoretical point of view. These features make it a tantalizing target for the search of potentially exotic emitters (and subdominant PBHs in particular). Indeed, the bounds on the PBH abundance from gamma-rays in this region \cite{Carr:2009jm} necessarily use the approach that we have termed {\it conservative} (i.e.\ no astrophysical background is assumed).

Returning now to PBHs as a dark matter candidate, i.e.\ for masses $\sim 10^{18}$~g, the fact that $\Phi_M$ decreases as $M$ increases (and the measured diffuse background increases at the corresponding, lower, energies)  a more accurate measurement and modeling of the soft X-ray diffuse background may  play, as we have argued, a crucial role towards a potential discovery or, at least, to strengthen current bounds substantially.  To conclude, we} remark that there exist other channels for PBH detection in {the mass window relevant for DM}. In particular, the energy injection associated to Hawking emission around or just after recombination can be probed with CMB data \cite{Poulin:2016anj}. Also, if the effect occurs during the reionization epoch it can be potentially detectable in the 21 cm absorption line of neutral Hydrogen \cite{Clark:2018ghm}. Moreover, the positron emission by PBHs of mass around $10^{17}$~g has been recently used to set a bound on $f$ using the keV line, which can improve over the gamma-ray bound, depending on the assumed Galactic density profile \cite{DeRocco:2019fjq}. All these channels are complementary and their further exploration could help to identify or rule out a significant population of PBHs in this promising mass window.

%

\vspace{0.5cm}

\mysections{Acknowledgments}
We thank M.A.\ S\'anchez-Conde for collaboration in the first stages of this project, M.\ Ajello for providing files with the X-ray data,  P.\ Serpico and M.\ Taoso for discussions and useful comments and suggestions on a draft version of this work, J.R.\ Espinosa and A.\ Urbano for discussions and A.\ Arbey and J.\ Auffinger for correspondence about BlackHawk. We thank M.\ Boudaud for insightful comments which helped to improve this paper. 
The work of GB is funded by a {\it Contrato de Atracci\'on de Talento (Modalidad 1) de la Comunidad de Madrid} (Spain), with number 2017-T1/TIC-5520, by {\it MINECO} (Spain) under contract FPA2016-78022-P and {\it MCIU} (Spain) through contract PGC2018-096646-A-I00.
DG has received financial support through the Postdoctoral Junior Leader Fellowship Programme from la Caixa Banking Foundation (grant n.\ LCF/BQ/LI18/11630014).
JCB is supported by an {\it Atracci\'on de Talento de la Comunidad de Madrid} (contract no.\ 2016-T1/TIC-1542). 
DG and JCB are additionally supported by {\it MINECO} (Spain) through the grant PGC2018-095161-B-I00 and Red Consolider MultiDark FPA2017-90566-REDC.
All authors are supported by the IFT UAM-CSIC Centro de Excelencia Severo Ochoa SEV-2016-0597 grant.

\bibliographystyle{hunsrtm}
\bibliography{PBHbiblio}

\begin{thebibliography}{10}

\bibitem{Hill:2018trh}
R.~Hill, K.~W. Masui, and D.~Scott.
\newblock {\em The Universe}, 72:663--688, 2018, 1802.03694.

\bibitem{ZelNov}
B.~Zel'dovich and I.~D. Novikov.
\newblock {\em Ya. Astron.Zh.}, 43,758.

\bibitem{Hawking:1971ei}
S.~Hawking.
\newblock {\em Mon. Not. Roy. Astron. Soc.}, 152:75, 1971.

\bibitem{Carr:2009jm}
B.~J. Carr, K.~Kohri, Y.~Sendouda, and J.~Yokoyama.
\newblock {\em Phys. Rev.}, D81:104019, 2010, 0912.5297.

\bibitem{Barnacka:2012bm}
A.~Barnacka, J.~F. Glicenstein, and R.~Moderski.
\newblock {\em Phys. Rev.}, D86:043001, 2012, 1204.2056.

\bibitem{Katz:2018zrn}
A.~Katz, J.~Kopp, S.~Sibiryakov, and W.~Xue.
\newblock {\em JCAP}, 1812:005, 2018, 1807.11495.

\bibitem{Graham:2015apa}
P.~W. Graham, S.~Rajendran, and J.~Varela.
\newblock {\em Phys. Rev.}, D92(6):063007, 2015, 1505.04444.

\bibitem{Capela:2013yf}
F.~Capela, M.~Pshirkov, and P.~Tinyakov.
\newblock {\em Phys. Rev. D}, 87(12):123524, 2013, 1301.4984.

\bibitem{Montero-Camacho:2019jte}
P.~Montero-Camacho, X.~Fang, G.~Vasquez, M.~Silva, and C.~M. Hirata.
\newblock 2019, 1906.05950.

\bibitem{Niikura:2017zjd}
H.~Niikura et~al.
\newblock {\em Nat. Astron.}, 3(6):524--534, 2019, 1701.02151.

\bibitem{Boudaud:2018hqb}
M.~Boudaud and M.~Cirelli.
\newblock {\em Phys. Rev. Lett.}, 122(4):041104, 2019, 1807.03075.

\bibitem{DeRocco:2019fjq}
W.~DeRocco and P.~W. Graham.
\newblock {\em Phys. Rev. Lett.}, 123(25):251102, 2019, 1906.07740.

\bibitem{Laha:2019ssq}
R.~Laha.
\newblock {\em Phys. Rev. Lett.}, 123(25):251101, 2019, 1906.09994.

\bibitem{Dasgupta:2019cae}
B.~Dasgupta, R.~Laha, and A.~Ray.
\newblock 12 2019, 1912.01014.

\bibitem{Poulin:2016anj}
V.~Poulin, J.~Lesgourgues, and P.~D. Serpico.
\newblock {\em JCAP}, 03:043, 2017, 1610.10051.

\bibitem{Stocker:2018avm}
P.~Stocker, M.~Kramer, J.~Lesgourgues, and V.~Poulin.
\newblock {\em JCAP}, 03:018, 2018, 1801.01871.

\bibitem{Carr:2016hva}
B.~Carr, K.~Kohri, Y.~Sendouda, and J.~Yokoyama.
\newblock {\em Phys. Rev. D}, 94(4):044029, 2016, 1604.05349.

\bibitem{Laha:2020ivk}
R.~Laha, J.~B. Mu\~noz, and T.~R. Slatyer.
\newblock 4 2020, 2004.00627.

\bibitem{Hawking:1974sw}
S.~W. Hawking.
\newblock {\em Commun. Math. Phys.}, 43:199--220, 1975.

\bibitem{Hawking:1974rv}
S.~W. Hawking.
\newblock {\em Nature}, 248:30--31, 1974.

\bibitem{Page:1976wx}
D.~N. Page and S.~W. Hawking.
\newblock {\em Astrophys. J.}, 206:1--7, 1976.

\bibitem{MacGibbon:1990zk}
J.~H. MacGibbon and B.~R. Webber.
\newblock {\em Phys. Rev.}, D41:3052--3079, 1990.

\bibitem{MacGibbon:1991vc}
J.~H. MacGibbon and B.~J. Carr.
\newblock {\em Astrophys. J.}, 371:447--469, 1991.

\bibitem{Strong:2004ry}
A.~W. Strong, I.~V. Moskalenko, and O.~Reimer.
\newblock {\em Astrophys. J.}, 613:956--961, 2004, astro-ph/0405441.

\bibitem{2010PhRvL.104j1101A}
A.~A. {Abdo et al.}
\newblock {\em \prl}, 104(10):101101, Mar 2010, 1002.3603.

\bibitem{Weidenspointner:2000aq}
G.~Weidenspointner et~al.
\newblock {\em AIP Conf. Proc.}, 510(1):581--585, 2000, astro-ph/0012332.

\bibitem{Arbey:2019vqx}
A.~Arbey, J.~Auffinger, and J.~Silk.
\newblock {\em Phys. Rev. D}, 101(2):023010, 2020, 1906.04750.

\bibitem{Aghanim:2018eyx}
N.~Aghanim et~al.
\newblock 2018, 1807.06209.

\bibitem{Arbey:2019mbc}
A.~Arbey and J.~Auffinger.
\newblock {\em Eur. Phys. J. C}, 79(8):693, 2019, 1905.04268.

\bibitem{Swift_paper}
{Scott D. Barthelmy et al.}
\newblock {\em Space Science Reviews}, 120(3-4):143--164, oct 2005.

\bibitem{MAXI_paper}
{Masaru Matsuoka et al.}
\newblock {\em Publications of the Astronomical Society of Japan},
  61(5):999--1010, oct 2009.

\bibitem{ASCA_paper}
Y.~Tanaka, H.~Inoue, and S.~S. Holt.
\newblock {\em Publications of the Astronomical Society of Japan}, 46:L37--L41,
  June 1994.

\bibitem{XMM_paper}
D.~H. Lumb.
\newblock {\em Optical Engineering}, 51(1):011009, feb 2012.

\bibitem{Chandra_paper}
{Martin C. Weisskopf et al.}
\newblock In J.~E. Truemper and B.~Aschenbach, editors, {\em X-Ray Optics,
  Instruments, and Missions {III}}. {SPIE}, jul 2000.

\bibitem{ROSAT_paper}
{I. M. McHardy et al.}
\newblock {\em Monthly Notices of the Royal Astronomical Society},
  295(3):641--671, apr 1998.

\bibitem{Ueda:2014tma}
Y.~Ueda, M.~Akiyama, G.~Hasinger, T.~Miyaji, and M.~G. Watson.
\newblock {\em Astrophys. J.}, 786:104, 2014, 1402.1836.

\bibitem{Urry:1995mg}
C.~M. Urry and P.~Padovani.
\newblock {\em Publ. Astron. Soc. Pac.}, 107:803, 1995, astro-ph/9506063.

\bibitem{DeAngelis:2017gra}
M.~Tavani et~al.
\newblock {\em JHEAp}, 19:1--106, 2018, 1711.01265.

\bibitem{Watanabe1997}
{K. Watanabe et al.}
\newblock In {\em {AIP} Conference Proceedings}. {AIP}, 1997.

\bibitem{Fukada1975}
{Y. Fukada et al.}
\newblock {\em Astrophysics and Space Science}, 32(1):L1--L5, jan 1975.

\bibitem{Gruber1999}
{D. E. Gruber et al.}
\newblock {\em The Astrophysical Journal}, 520(1):124--129, jul 1999.

\bibitem{Kinzer1997}
{R. L. Kinzer et al.}
\newblock {\em The Astrophysical Journal}, 475(1):361--372, jan 1997.

\bibitem{Harada:2017fjm}
T.~Harada, C.-M. Yoo, K.~Kohri, and K.-I. Nakao.
\newblock {\em Phys. Rev. D}, 96(8):083517, 2017, 1707.03595.
\newblock [Erratum: Phys.Rev.D 99, 069904 (2019)].

\bibitem{McEnery:2019tcm}
R.~Caputo et~al.
\newblock 1907.07558.

\bibitem{Hunter:2013wla}
S.~D. Hunter et~al.
\newblock {\em Astropart. Phys.}, 59:18--28, 2014, 1311.2059.

\bibitem{Tomsick:2019wvo}
J.~A. Tomsick et~al.
\newblock 1908.04334.

\bibitem{Tanimori:2015wma}
T.~Tanimori et~al.
\newblock {\em Astrophys. J.}, 810(1):28, 2015, 1507.03850.

\bibitem{Matsumoto:2018mba}
H.~Matsumoto et~al.
\newblock {\em J. Astron. Telesc. Instrum. Syst.}, 4(1):011212, 2018,
  1803.00242.

\bibitem{2009ApJ...699..603A}
M.~{Ajello}, L.~{Costamante}, R.~M. {Sambruna}, N.~{Gehrels}, J.~{Chiang},
  A.~{Rau}, A.~{Escala}, J.~{Greiner}, J.~{Tueller}, J.~V. {Wall}, and R.~F.
  {Mushotzky}.
\newblock {\em \apj}, 699(1):603--625, July 2009, 0905.0472.

\bibitem{Ivanov:1994pa}
P.~Ivanov, P.~Naselsky, and I.~Novikov.
\newblock {\em Phys. Rev.}, D50:7173--7178, 1994.

\bibitem{Garcia-Bellido:2017mdw}
J.~Garcia-Bellido and E.~Ruiz~Morales.
\newblock {\em Phys. Dark Univ.}, 18:47--54, 2017, 1702.03901.

\bibitem{Kannike:2017bxn}
K.~Kannike, L.~Marzola, M.~Raidal, and H.~Veermäe.
\newblock {\em JCAP}, 1709(09):020, 2017, 1705.06225.

\bibitem{Ballesteros:2017fsr}
G.~Ballesteros and M.~Taoso.
\newblock {\em Phys. Rev.}, D97(2):023501, 2018, 1709.05565.

\bibitem{Cicoli:2018asa}
M.~Cicoli, V.~A. Diaz, and F.~G. Pedro.
\newblock {\em JCAP}, 1806(06):034, 2018, 1803.02837.

\bibitem{Dalianis:2018frf}
I.~Dalianis, A.~Kehagias, and G.~Tringas.
\newblock {\em JCAP}, 1901:037, 2019, 1805.09483.

\bibitem{Ballesteros:2019hus}
G.~Ballesteros, J.~Rey, and F.~Rompineve.
\newblock {\em JCAP}, 06:014, 2020, 1912.01638.

\bibitem{Ballesteros:2020qam}
G.~Ballesteros, J.~Rey, M.~Taoso, and A.~Urbano.
\newblock 2001.08220.

\bibitem{Ajello:2008xb}
M.~Ajello et~al.
\newblock {\em Astrophys. J.}, 689:666, 2008, 0808.3377.

\bibitem{Clark:2018ghm}
S.~Clark, B.~Dutta, Y.~Gao, Y.-Z. Ma, and L.~E. Strigari.
\newblock {\em Phys. Rev.}, D98(4):043006, 2018, 1803.09390.

\end{thebibliography}

\end{document}